\documentclass[epj,referee]{svjour}
\usepackage{graphics}
\usepackage{graphics}
\begin{document}
\sloppy
\title
{Studies of bosons in optical lattices in a harmonic potential}
\author{R. Ramakumar\inst{1}, A. N. Das\inst{2}, and S. Sil\inst{3}}
\institute{Department of Physics and Astrophysics, University of Delhi,
Delhi-110007, India \and Condensed Matter Physics Group, 
Saha Institute of Nuclear Physics
1/AF Bidhannagar, Kolkata-700064, India 
\and Department of Physics, Visva Bharati, 
Santiniketan-731235, India} 
\date{18 January 2007}
\abstract
{  We present a theoretical study of bose condensation and specific heat 
of non-interacting bosons in finite lattices in harmonic potentials in
one, two, and three dimensions. We numerically diagonalize
the Hamiltonian to obtain the energy levels of the systems.
Using the energy levels thus obtained, we investigate the
temperature dependence, dimensionality effects, lattice size
dependence, and evolution to the bulk limit of the condensate
fraction and the specific heat. Some preliminary results on
the specific heat of fermions in optical lattices are also presented.
The results obtained are contextualized within the current experimental and
theoretical scenario.
\PACS{03.75.Lm, 03.75.Nt, 03.75.Hh}
} 
\maketitle
\section{Introduction}
\label{sec1}
Bosons in optical lattices provides a microscopic 
laboratory for exploration of the properties of many-boson 
systems with unprecedented control on
number of bosons per site, boson kinetic energy, and boson-boson
interaction strength \cite{greiner,stof,paredes,kohl,jak}.
Experimentalists have explored \cite{greiner,stof,paredes,kohl} quantum phase
transitions, excitation spectra, condensate fraction
as a function of boson-boson interaction strength, and
properties of boson Tonks gas in one, two, and three dimensional
optical lattices. Theoretical studies \cite{jak,oost,zwer,rey,pupilo,wessel,kleinert,ho,pollet,santos,sengupta,damle,pandit,scarola} 
of bosons in optical lattices thus far mainly
concentrated on quantum phase transitions between
various possible phases such as Mott insulator phase,
density-wave phase, and bose-condensed phase.
Finite temperature properties of bosons in the combined 
optical lattice potential and the confining harmonic 
potential have not received much attention except in
some studies \cite{dicker,giampaolo} where, again, the focus is 
on Mott transition.
\par
In this paper, we present a theoretical study of bosons in
optical lattices in a harmonic potential in one, two, and
three dimensional optical lattices. In particular, we
study the temperature dependence of the condensate fraction 
and specific heat. This study is useful for the following reasons.
After the discovery \cite{cornell,davis,bradley} of Bose-Einstein
condensation in harmonic traps
(for reviews see Refs. \cite{pita,leg,ketterle,pethick}), 
one of the important studies  done was to measure the temperature
dependence of the condensate fraction and its comparison with theory,
and such measurements for bosons in optical lattices
can be expected in near future. 
Thermodynamic properties of bosons in optical lattices are
of considerable interest in their own right. Though no measurement
of specific heat of bosons in a harmonic potential and
of bosons in a lattice with harmonic potential
have been reported thus far, such measurements have already been
reported \cite{kinast} for fermions in harmonic potential, and it
is not unreasonable to
expect such measurements for bosons as well. Furthermore,
experiments are performed on bosons in {\em finite} lattices,
and so it would be useful to understand the properties
of bosons in finite lattices to see finite size effects
and  their evolution towards the bulk limit. The motivations
and relevance of the studies presented in this paper
are, hopefully, clear to the reader from the foregoing.      
\par
As stated, we present studies of lattice bosons in harmonic traps.
We have considered non-interacting  bosons in a periodic lattice 
in one-dimension (1d), a square lattice in two-dimensions (2d),
and a cubic lattice in three-dimensions (3d). 
The bosons are considered to be moving in the combined lattice and
harmonic potentials.  We  numerically diagonalize 
the Hamiltonian matrix for bosons in finite lattices in
a harmonic potential, and concentrate on
calculations of the temperature
dependence of ground state occupancy, specific-heat, finite-size
effects, and dimensionality effects. 
This work is presented in the
Sections 2 and 3. In Section 4, some preliminary results on our
recent work on fermions in optical lattices in a harmonic potential is
presented, and conclusions are given 
in Section 5.
\section{Model and method}
\label{sec2}
In this Section, we will describe the model and the method
followed. Consider non-interacting bosons under the combined
influence of a  periodic optical lattice potential and  an overall 
harmonic confining potential. The Hamiltonian of this system is,
\begin{equation}
H=-t \sum_{<ij>}\left(c^{\dag}_{i}c_{j}+c^{\dag}_{j}c_{i}\right)
  +K\sum_{i}r^{2}_{i}n_{i}-\mu\sum_{i}n_{i}\,,
\end{equation}
where $t$ is the kinetic energy gain when a boson hop
from site $i$ to its nearest neighbor site $j$ in the optical lattice,
$c^{\dag}_{i}$ is the boson creation operator, 
$K$ ($>\,0$) the strength of the harmonic confining 
potential, ${\mathbf r}_{i}$ the locator of the site $i$ with
respect to the origin which is at the
center of the lattice, $n_{i}=c^{\dag}_{i}c_{i}$
the boson number operator, and $\mu$ the chemical potential.  
For bosons  under the influence
of only the harmonic potential bose condensation is 
by now well understood \cite{pita,leg,ketterle,pethick}.
Our aim is to consider the case of finite $t$ and finite $K$.
After writing the Hamiltonian (Eq.1) in the single particle site basis,
we numerically diagonalize it for various values of $K$ and 
for various lattice sizes in 1d, 2d, and 3d. 
The energy levels thus obtained are used in calculations of the 
ground state occupancy and specific heat as a function of
temperature.  We will see that any non-zero value of $K$
leads to substantial changes in both these properties.
We have calculated the energy levels and the bosonic properties 
using open boundary conditions. 
It may be noted here that for the 1d case, the eigenfunctions
are eigenenergies may be exactly obtained in terms of the Mathieu functions
as shown by Rey and collaborators \cite{rey1}. They have given approximate 
analytical expressions for the eigenenergies  and eigenfunctions. 
We compared our eigenenergies
for 1d systems with their approximate result (Eq. 15 in their
paper) in the tunneling dominated regime and found that their eigenenergies
are very close to our exact results for large values of $t/k$. 
\par
The ground state occupancy and the specific heat are calculated
in the following way. The population $N(E_{i})$ of a state
with energy $E_{i}$ is given by the Bose distribution
\begin{eqnarray}
N(E_{i})&=&\frac{1}{e^{\beta\,(E_{i}-\mu)}-1}\, , 
\end{eqnarray}
where $\beta\,=\,1/k_{B}T$ and $k_{B}$ is the Boltzmann 
constant. The chemical potential is
determined from the number equation
\begin{equation}
N\,=\,\sum_{i=0}^{i_{m}}N(E_{i})\, ,
\end{equation}
where $E_{i}$ with $i\,$= 0, 1, 2....., $i_{m}$ denotes 
the energy levels from the lowest to the highest level of
a boson in a finite lattice in the presence of the harmonic
confining potential. 
For a given number of bosons, at any temperature, the
chemical potential and the population in the lowest level ($N_{0}$) 
are determined using Eqs. (2) and (3).
\par
The total energy of the system of bosons is given by
\begin{equation}
E_{tot}=\sum_{i=0}^{i_{m}}N(E_{i})E_{i}.
\end{equation}
The temperature derivative of $E_{tot}$ gives 
the specific heat
\begin{equation}
C_{v}=\frac{1}{k_{B}T^{2}}
      \left[I_{3}-\frac{I_{2}I_{1}}{I_{\circ}}\right]\, ,
\end{equation}
where,
\begin{eqnarray}
I_{\circ}&=&\sum_{i=0}^{i_{m}} F(E_{i}-\mu)\, , \\
I_{1}&=&\sum_{i=0}^{i_{m}} E_{i}\,F(E_{i}-\mu)\, , \\
I_{2}&=&\sum_{i=0}^{i_{m}} (E_{i}-\mu)\,F(E_{i}-\mu)\, , \\
I_{3}&=&\sum_{i=0}^{i_{m}} E_{i} (E_{i}-\mu)\,F(E_{i}-\mu)\, .
\end{eqnarray}
In the above equations
\begin{equation}
F(E_{i}-\mu) = \frac{e^{\beta(E_{i}-\mu)}}
{\left[e^{\beta(E_{i}-\mu)}-1\right]^{2}}\,.
\end{equation}
\section{Results and Discussions on bosons}
\label{sec3}
Before presenting the results of our calculations, some general 
remarks are in order. In the absence of a harmonic 
confining potential, the bosons in a
periodic lattice undergo a bose condensation phase transition only in 3d where
the boson Density Of States (DOS) in the thermodynamic limit vanishes at the
bottom of the band. The phase transition is not possible for bosons
in 1d or 2d  periodic lattices since the DOS at the bottom
of the band either diverges or remains finite, and the  condition
$N_{0} = 0$ and $\mu=E_{0}$ is not satisfied at any temperature.
Here, $E_0$ is the energy of the lowest level or bottom of a band.
For free bosons in a harmonic confining potential, the phase transition
is possible in 2d but not in 1d \cite{bagnato}. For finite size
systems and for finite number of particles there is no true phase
transition using the above criteria. However, at sufficiently low
temperatures, a macroscopic number of particles will occupy the
ground state and thus bose condensation will occur \cite{vandrut}.
With these general remarks, we go
to our results.
\subsection{One Dimensional case}
In Fig. 1, the variation of the condensate fraction ($N_{0}/N$)
with temperature is shown for 1000 bosons in a 1d lattice 
(of lattice constant $a$) for $k$ = $Ka^{2}$ = 0.01 (in the
energy scale of t = 1) for different lattice sizes (N$_{l}$).
The occupancy in the lowest level  is considerable even at high 
temperatures for smaller size systems and it
decreases with increasing size. For $k$  = 0.01, no size effect is seen
for N$_{l} \geq$ 400. We have presented a curve for N$_{l}$ = 1000
which may be considered as an infinite size result for this value
of $k$.
\par
In Fig 2. we have shown the temperature dependence of the condensate
fraction ($N_0/N$) for a one dimensional lattice of size $N_l = 1000$ in
presence of harmonic potential of strength $k= 0.01$ for different numbers
of bosons. The $N_{0}/N$ is plotted against $T/T_{0}$.
Here, $T_{0}$ is determined by setting $N_{0}=0$ and $\mu =E_{0}$ in the number
equation (Eq. 3) to obtain 
\begin{equation}
N = \sum_{i=1}^{i_{m}}\frac{1}{e^{(E_{i}-E_{0})/k_{B}T_{0}}-1}\, .
\end{equation}
Note that in the thermodynamic limit and when the DOS vanishes
at the bottom of the boson energy band, $T_{0}$ gives the phase
transition temperature. For finite size lattices $T_{0}$
may be considered as the {\em condensation temperature}. 
In the absence of a harmonic trap, $N_{0}/N$ is appreciable
for bosons in a 1d lattice even at $T\,=\,2T_{0}$ (dash-dot curve
in Fig. 2). It decreases rapidly with the application of the harmonic 
potential and also decreases with increasing number of bosons 
for a fixed value of $k$. For a bose gas in a harmonic
trap (dashed curve in Fig. 2), $N_{0}/N$ almost vanishes at
high temperatures. In Fig. 3, the variation of $T_0$ with number 
of bosons (N) is shown for different values of the harmonic 
potential. The $T_0$ increases 
almost linearly with N.  For a fixed N, it increases with increasing
strength of the harmonic potential. 
\par

In Fig. 4, we have plotted the specific heat per particle against $T/T_0$
for different values of $k$ for $N_l=1000$ and $N\,=\,600$. It is seen that the 
specific heat decreases with increasing value of $k$. For $k$ = 0.001,
a broad peak is seen in the specific heat, which is a signature of
the finite size of the lattice.
For such a small value 
of $k$, even a size of 1000 sites is not enough to confine the
bosons within the harmonic trap (i. e., some of the bosons
may occupy  the boundary sites), hence finite size effects
would show up. As the strength of the harmonic potential increases,
the number bosons at  high-energy sites at the edges of
the lattice is insignificant and the finite size effect disappears. 
We have also shown $C_v$ for bosons under
the influence of only the harmonic potential. 
One notices at once that while the dependence of $C_v$ on the
scaled temperature ($T/T_0$) for free bosons in a
harmonic potential is independent of $k$, the same is
not true for lattice bosons. It is found that,
switching on the lattice potential substantially reduces
the $C_v$. We also note that reducing the strength of
harmonic potential for the lattice bosons drives the system towards
the limit of free bosons in a harmonic potential, which is counter 
intuitive to what one might naively expect.
In the very low temperature range,  
all the curves are very close to each other. This 
can be understood if we look at the single
particle spectrum of a large lattice. In this case, the
low energy part of the band is approximately parabolic, and
so one can transform the problem to a free particle in a
harmonic potential.  Hence one expects the properties of
lattice bosons in a harmonic trap will be close to that
of free bosons in  harmonic trap at sufficiently low
temperatures.  
\subsection{Two Dimensional case}
In this section we present our results on bosons in two dimensional
optical lattices in a 2d harmonic potential. These results
are presented in Figs. 5-10. Fig. 5 shows the variation of the 
condensate fraction with temperature for different lattice 
sizes for $k$ = 0 and 0.01, respectively. A macroscopic 
condensation occurs at a lower temperature for larger size of the system. 
The dashed lines in Fig. 5 exhibit the effect of harmonic
potential on lattice bosons in 2d. In the presence of a harmonic
confining potential,  $N_0/N$ versus $T$ curves become almost 
size independent for lattice sizes of $80\times 80$ and larger,
unlike the case of $k$ = 0 (solid lines) where the condensation
temperature decreases rapidly with the size of the system.
For largest lattice size shown, in presence of the harmonic 
potential our results may correspond 
to the results in the thermodynamic limit.
It may be mentioned that for larger number of bosons
a larger lattice size is needed to reach the infinite size limit.
The $T_0$ is clearly seen to be considerably increased 
with the introduction of the harmonic potential. 
We also note that the condensate fraction grows faster
with decreasing temperature in 2d as compared to that in 1d
(see Figs. 1 and 2). In Fig. 6, we have shown the variation 
of $T_{0}$ with boson number. The figure shows that, for a fixed lattice 
size and $N$, $T_{0}$ increases
with increasing strength of the harmonic confinement potential. It also shows
that the bosons under the combined effect of lattice and
harmonic potentials have a higher transition temperature compared to
bosons with either of the potentials absent. In comparison with 1d,
$T_{0}$ is found to increase rapidly for small $N$ and a
monotonic increase for larger $N$, clearly showing a
dimensionality effect.
\par
In Fig. 7, we have exhibited the effect of lattice size on the specific heat
of bosons for a fixed $N$ and $k$. 
For small lattice sizes, we find a flat region in $C_v$ for
temperatures below $T_{0}$. With increasing lattice size the flat
region evolves into a peak which becomes sharper as the size of the lattice 
approaches the bulk limit. If one compares the $C_v$ results for
$80 \times 80$, $100 \times 100$, $150 \times 150$ lattices, 
one finds that in the low temperature range the value of $C_v$ is
same for all the three curves, while in the high temperature range 
$C_v$ is smaller for smaller size systems. The latter is a signature of
the size effect.
For higher temperatures more and more 
bosons would occupy the boundary sites of the lattice for a given
value of $k$ leading to appreciable finite size effects. 
Given the great control experimentalists have demonstrated,
these effects may be within the reach of experimental observability
in near future.
\par
In order to get an understanding of the effect of harmonic potential
strength on the specific heat of lattice bosons, we have shown in
Fig. 8, $C_v$ as a function of $T/T_0$ for various values of $k$.
This figure shows that the $C_v$ curves for different $k$ values
intercept at a temperature $T/T_{0} \approx 1.1$ for the boson
number ($N$ = 3840) considered. Below this
temperature $C_v$ is lower for higher $k$, while above it increasing
$k$ increases $C_v$. This is in contrast to the behavior of
bosons in 1d (Fig. 4) where increasing $k$ decreases $C_v$ except
at very low temperatures. The peak in $C_v$ curve is seen to decrease
with increasing harmonic confinement. 
The behavior of $C_v$ shown in Fig 8 results from an 
interplay of the effects of the harmonic potential and  
finite size of the lattice.
The cross-over effect for $T/T_{0} \geq 1.1$ occurs because
in this temperature range bosons are delocalized over the entire
lattice so that they sense the existence of boundaries. This
is clear from Fig. 9 in which $C_v$ is shown for a larger
lattice ($150 \times 150$) and one notices that the crossing
has disappeared in the temperature range shown. 
\par
In order to get an insight into the dimensionality effects,
we calculated the one boson density of states for a lattice
boson in a harmonic potential. The normalized DOS for
the 2d case is shown in Fig. 10. In the absence of the harmonic
potential, the DOS has a van-Hove singularity for a 2d square
lattice. Any finite harmonic potential leads
to drastic changes in DOS by wiping out the van-Hove
singularity and flattening the DOS in an energy range which
increases with increasing harmonic potential strength ($k$).
This range also increases with increasing size of the lattice
(not shown in the figure). The DOS at the bottom of the band
vanishes as the size of the lattice approaches the infinite limit,
and in the low energy range it shows almost linear increase
with energy before taking the constant value. The  DOS for bosons
in a 2d lattice in presence of a harmonic trap may thus be 
approximately given by 
\begin{eqnarray}
\rho(E)&=&\gamma E~~~{\rm for}~~~ 0\leq E\leq E_{1} \nonumber \\
         &=&\gamma E_{1}~~~{\rm for}~~~ E_{1}\leq E\leq E_{2}\, ,
\end{eqnarray} 
where all the energies are measured from the bottom of the energy
spectrum. It is found that $\gamma$ is strongly $k$ dependent and decreases
with increasing $k$, while $E_1$ is almost $k$ independent. 
For free bosons in a 2d harmonic trap, the DOS has
a linear dependence on $E$ in the entire energy range
($0 \leq E \leq \infty$). 
For $\rho(E) \propto E^{\alpha-1}$, the specific heat per boson 
for $T \leq T_0$ is given by\cite{pethick}
\begin{equation}
\frac{C_v}{Nk_B}=\alpha(\alpha+1)\frac{\zeta(\alpha+1)}{\zeta(\alpha)}
{\left(\frac{T}{T_0}\right)}^{\alpha}\,,
\end{equation}
where $\zeta(\alpha)=\sum_{n=1}^{\infty}(1/n^{\alpha})$ is the
Riemann zeta function. Since for free bosons in a harmonic
trap $\alpha=2$, $C_v$ shows a $(T/T_0)^{2}$ behavior for 
$T\leq T_0$. This feature is seen in our calculated results
(the dotted curve in Fig. 9). 
For the lattice bosons in a harmonic trap in the low
temperature region, only the low energy linear part of
of the DOS is occupied. Hence $C_v$ shows behavior
similar to that of free bosons in a harmonic trap.
When the temperature is appreciable so that the flat
portion of the DOS is also occupied, the effective value of
$\alpha$ decreases. This results in a slower increase
of $C_v$ with $T/T_0$ and lower value of $C_v$ compared
to the free bosons case. These features are prominent
in Fig. 9 for higher values of $k$ which have higher
$T_0$ values.
\subsection{Three Dimensional case}
In Fig. 11-13, we have presented our results for bosons in
3d cubic lattices in a harmonic potential. The
temperature dependence of the condensate fraction (Fig. 11)
and the boson number dependence of the condensation 
temperature (Fig. 12) shows variations similar
to those found for 2d systems. We find that $T_0$ increases with increasing
harmonic potential strength as well as with the total number of bosons in the
lattice. 
In Fig. 13, we have shown the effect of harmonic potential
strength on the specific heat of 3d lattice bosons.
The peak value in $C_v$ at $T_/T_0 \approx 1$ is lower for
higher values of $k$. For the $50\times 50 \times 50$ lattice,
$C_v$ is lower for lower values of $k$ at high temperatures ($T/T_0 > 1.3$).
This results from the finite size of the lattice, as mentioned previously. 
In comparison with a 2d system, the harmonic potential
strength  has a lesser effect in 3d (see Figs. 9 and 13).
In Fig. 13, we have also plotted the $C_v$ for free bosons
in a harmonic trap. The $C_v$ shows a $(T/T_0)^3$
dependence for $T/T_0 \leq 1$, which is a known result \cite{pethick}.
The figure shows that the $C_v$ for lattice bosons in
a harmonic trap has the same temperature dependence 
in the low temperature range. 
We have calculated the DOS for the bosons in the 3d lattice 
(of size $50\times 50 \times 50$) confined 
in a harmonic trap and found that it  has an $E^2$ dependence in the low
energy range. This accounts for the $(T/T_0)^3$ behavior
of the $C_v$ for the lattice bosons in a harmonic trap
in the low temperature range.

 In the work presented on bosons, the boson-boson interaction
is neglected.
We believe that our results would approximately hold when interaction 
energy ($U$) is much smaller than the hopping energy ($t$). 
In an optical lattice, the interaction energy ($U$) depends
on the ratio between the depth of the optical potential to the recoil energy.
A weakly interacting regime may be obtained by adjusting the value
of the potential depth to a low value \cite{greiner,schori}.
In the weak interaction regime, the interaction induced depletion effects 
may not be significant.  
\section{Results on Fermions}
\label{sec4}
Due to recent interest \cite{kohl2,kohl3} in fermions 
in optical lattices, we have studied
single (spin) component fermi gas in optical lattices with harmonic 
confinement in 1d, 2d, and 3d. 
Another interest in studying this fermionic system is that 
the bosons in the strongly interacting regime behave similar 
to fermions to avoid double occupancy and minimize inter-particle 
repulsion energy  
and energy spectrum of strongly interacting bosons in 1d is very 
close to that of the non-interacting fermions \cite{rey1}. 

Our results for the specific heat
for different values of the trapping strength and for different dimensions
are presented in Figs. 14-16. The specific heat curves show a linear 
variation at low temperatures, which is a characteristic of a degenerate
fermi gas. At high temperatures, the specific heat shows a more or less 
flat behavior.
The specific heat increases with decreasing $k$ values except
at low temperatures. Similar results are also noted for bosons 
(see Figs. 4 and 9).
With increasing $k$ value, the specific heat for fermions approaches 
the corresponding value for the $k=0$ case.  

We have determined the gradients of the specific heat curves 
at low temperatures. For 3d, this value is 9.76 for $k=0.01$ and
4.98 for $k=0$. The former is close to the corresponding value ($\pi^2$) 
for free fermions in a harmonic trap while the latter is close 
to that ($\pi^{2}/2$) of free fermions.
For the 2-d case, this slope is 7.2 for $k=0.01$ and 
3.37 for $k=0$, while for 1-d it is 3.95 for $k=0.0001$ and 1.69 for $k=0$.
We find that in all dimensions, the values of the gradient of the
specific heat curves at low temperatures
for small ($k/zt$) (where, $z$ is the coordination number of a lattice), 
are close to those of free fermions in a harmonic trap. The reason behind
it lies in the fact that the single particle density of states for 
small $k$ values in an optical lattice shows the same energy dependence 
as that of the free fermions (bosons) in a harmonic
trap at low energies, as discussed in previous sections.  
\section{Conclusions}
\label{sec5}
In this paper, we have presented a study of non-interacting bosons
under the influence of combined optical lattice 
and harmonic potentials in one, two, and three
dimensions. The condensate fraction in 1d shows a faster
reduction with increasing temperature in the low temperature
range compared to that in 2d and 3d. The condensation
temperature in 1d increases linearly with increasing number 
of lattice bosons. In comparison, $T_0$ in 2d and 3d shows a fast growth
for small number of bosons and a monotonic increase thereafter.
It is found that, for a given number of bosons,
the condensation temperature is higher for bosons in the 
combined harmonic and optical potentials compared to the cases 
in which either of the potentials is absent.
  
The specific heat results show several interesting features. 
In 1d, specific heat of lattice bosons in harmonic
potential is found to show a very slow growth with temperature,
except in the low temperature range where the growth is relatively faster. 
In 2d and 3d, $C_v$ of lattice bosons in a harmonic trap
has a peak at the condensation temperature. 
When the lattice size effects are important in $C_v$, we find that
the $C_v$ per boson is lower for smaller lattice sizes except at
low temperatures. In 2d, flat regions are observed in $C_v$ 
for a substantial temperature range below the condensation temperature for
relatively smaller lattices. 
When the lattice size effects are negligible, 
we find that the lattice bosons in a harmonic
trap has a considerably reduced specific heat compared to that
for free bosons in a harmonic trap for temperatures around and above $T_0$.
In all dimensions, the low temperature $C_v$ is very 
close to that of free bosons in a harmonic trap. 
The specific heat of the lattice 
bosons is strongly dependent on the strength of the harmonic potential
in contrast to that of free bosons in a harmonic potential. 
In all dimensions for large size systems when the lattice size
effects are negligible, the $C_v$ is lower for higher values
of $k$ around and above $T_0$. However, 
the $C_v$ curves for different $k$ values cross each other above $T_0$ 
when finite size effects are present.  
Considering recent interest in fermions in optical lattices,
we presented some preliminary results on the specific heat
of fermions in optical lattices in a harmonic potential in
1d, 2d, and 3d. The specific heat is found to show a linear
temperature dependence at low temperatures and a flat behavior
at high temperatures. With increasing strength of the harmonic
potential, the specific heat is found to approach
the pure lattice fermion results.
The temperature dependence of 
the specific heat of lattice bosons and fermions
in a harmonic trap is governed by a complex interplay of the
delocalizing effects of the boson kinetic energy, confining
effect of the harmonic potential, and the thermal energy.
\section*{Acknowledgments}
RRK thanks Professor Bikash Sinha, Director, SINP and 
Professor Bikas Chakrabarti, Head, TCMP Division, SINP 
for hospitality at SINP.
{}
\newpage
\begin{figure}
\resizebox*{3.1in}{2.5in}{\rotatebox{270}{\includegraphics{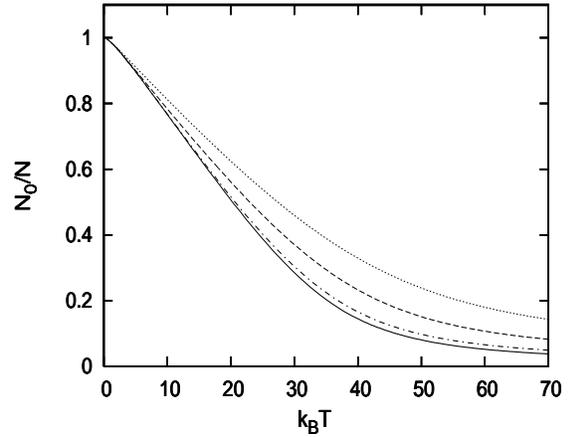}}}
\caption[]{
The variation of condensate fraction with temperature
for 1000 bosons
in  1d finite lattices in a harmonic potential with $k$ = 0.01.
The size of the lattices are:
N$_{l}$ = 30 (dots), 50 (dashes), 100 (dash-dot), and 1000 (solid line). 
k$_{B}$T is expressed in energy units of t = 1.} 
\end{figure}
\begin{figure}
\resizebox*{3.1in}{2.5in}{\rotatebox{270}{\includegraphics{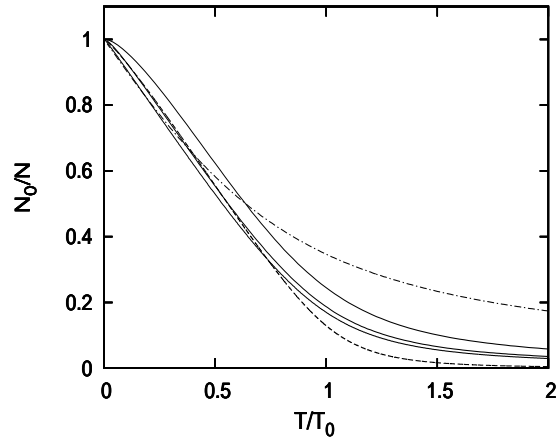}}}
\caption[]{
The variation of condensate fraction with temperature. Solid lines
are for bosons
in a 1d finite lattice of size 1000 in a harmonic potential with $k$ = 0.01
for boson numbers $N\,=\,100$ (top),
$1000$ (middle), $10000$ (bottom). The dashed line is for 10000 free bosons
in  a harmonic trap with $k$ = 0.01.
The dash-dot line is for 10000 bosons in a 1d lattice of 1000 sites. 
In this and other figures $T_{0}$ is the condensation
temperature (see text).}
\end{figure}
\begin{figure}
\resizebox*{3.1in}{2.5in}{\rotatebox{270}{\includegraphics{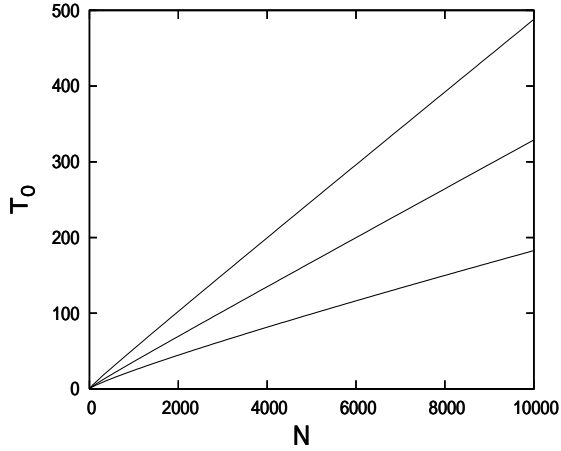}}}
\caption[]{The variation of $T_{0}$ (in energy units of t=1)
 with $N$ for bosons in a 
1d lattice of size 1000 in a 1d harmonic 
trap with $k$ = 0.02 (top), $0.01$ (middle),
and for free bosons in a harmonic trap for $k$ = 0.01 (bottom).
}
\end{figure}
\begin{figure}
\resizebox*{3.1in}{2.5in}{\rotatebox{270}{\includegraphics{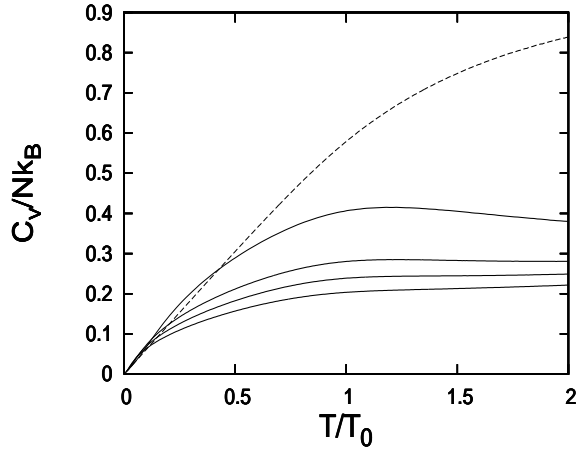}}}
\caption[]{
Temperature dependence of the specific heat per particle for 600 bosons. 
The dashed line is for free bosons in a harmonic potential. 
Solid lines are for bosons
in a 1d optical lattice of size 1000 in a harmonic potential 
for (from top to bottom): $k$ = 0.001(top), 0.005, 0.01, and 0.02 (bottom).} 
\end{figure}
\begin{figure}
\resizebox*{3.1in}{2.5in}{\rotatebox{270}{\includegraphics{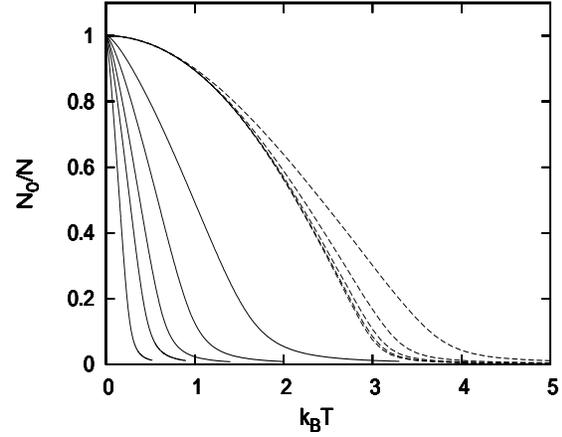}}}
\caption[]{Size effect in 2d for N = 540. The dashed curves are
for $k$ = 0.01 and lattice sizes: $30 \times 30$ (top), $40 \times 40$,
$50 \times 50$, $60 \times 60$, and $80 \times 80$ (bottom). Solid
curves are for lattice bosons without harmonic potential for
lattice sizes: $30 \times 30$ (top), $40 \times 40$, $50 \times 50$,
$60\times60$, and $80 \times 80$ (bottom). $k_{B}$T is measured in 
energy units of t = 1.}
\end{figure}
\begin{figure}
\resizebox*{3.1in}{2.5in}{\rotatebox{270}{\includegraphics{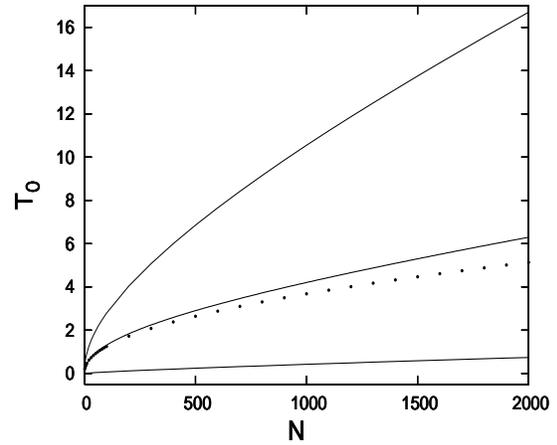}}}
\caption[]{The variation of T$_{0}$ (in energy units of t =1)
with number of bosons. Solid lines are  for
an $80 \times 80$ lattice and with $k$ = 0.05 (top), 0.01 (middle),
and $k$ = 0 (bottom). The dots are for bosons with only the
harmonic trap potential with $k$ = 0.01.
}
\end{figure}
\begin{figure}
\resizebox*{3.1in}{2.5in}{\rotatebox{270}{\includegraphics{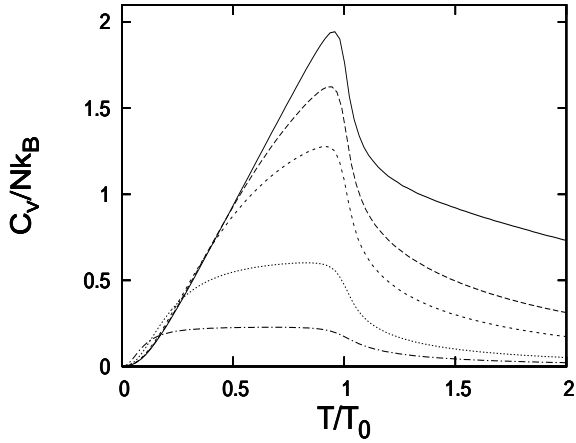}}}
\caption[]{Lattice size effect on specific heat in 2d for
3840 bosons in a harmonic trap with $k$ = 0.01. The curves are
for lattice sizes:
 150 $\times$ 150 (solid line),
 100 $\times$ 100 (long dashes line),
 80 $\times$ 80 (short dashed line),
 50 $\times$ 50 (dotted line), and
 30 $\times$ 30 (dash-dot line).
}
\end{figure}
\begin{figure}
\resizebox*{3.1in}{2.5in}{\rotatebox{270}{\includegraphics{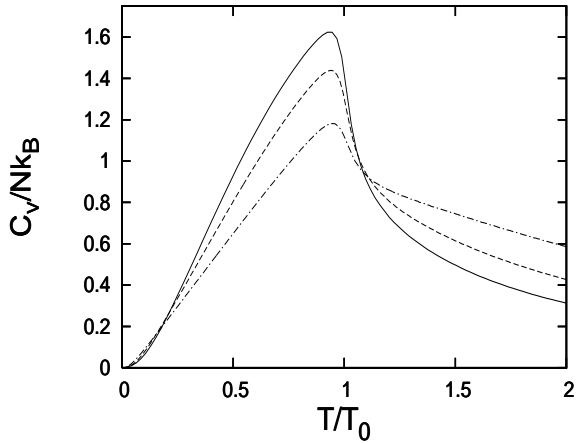}}}
\caption[]{
Specific heat per boson in 2d for 3840 bosons in a $100\times 100$ lattice
in a harmonic potential with $k$ = 0.01 (solid line), $k$ = 0.02 (dash line),
and $k$ = 0.05 (dash-dot line).}
\end{figure}
\begin{figure}
\resizebox*{3.1in}{2.5in}{\rotatebox{270}{\includegraphics{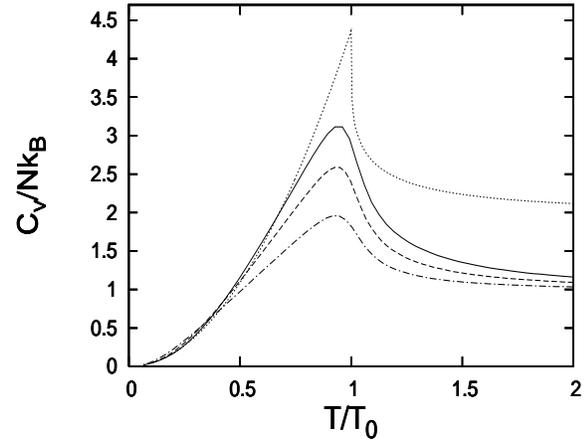}}}
\caption[]{
Specific heat per boson in 2d for 540 bosons in a $150\times 150$ lattice
in a harmonic potential with $k$ = 0.01 (solid line), $k$ = 0.02 (dashed
line),
and $k$ = 0.05 (dash-dot line). The dotted curve (top) is for bosons
in 2d harmonic trap and is independent of k.}
\end{figure}
\begin{figure}
\resizebox*{3.1in}{2.5in}{\rotatebox{270}{\includegraphics{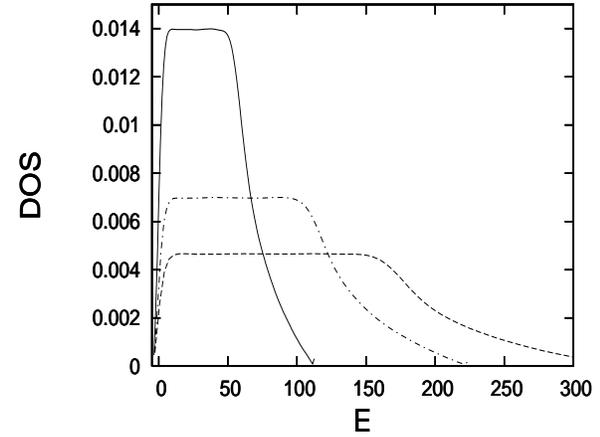}}}
\caption[]{
Normalized DOS per site versus energy for an $100\times100$ lattice for
$k$ =0.01 (solid line) , 0.02 (dash-dot line), and 0.03 (dashed line).
}
\end{figure}
\begin{figure}
\resizebox*{3.1in}{2.5in}{\rotatebox{270}{\includegraphics{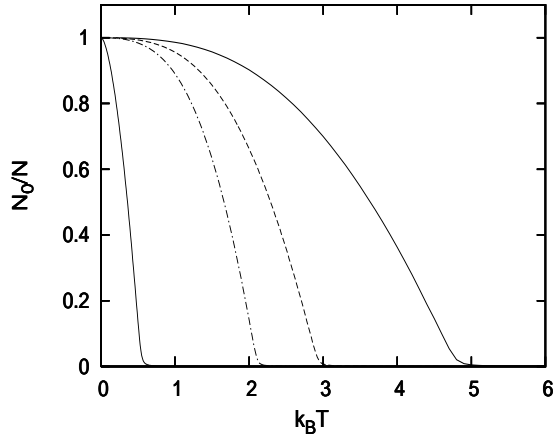}}}
\caption[]{Temperature dependence of condensate fraction for 2500 bosons
in a cubic lattice with $50\times 50 \times 50$ lattice sites and
harmonic potential with $k$ = 0.05 (top), $k$ = 0.02 (dashed),
$k$ = 0.01 (dash-dot), and $k$ = 0 (bottom).}
\end{figure}
\begin{figure}
\resizebox*{3.1in}{2.5in}{\rotatebox{270}{\includegraphics{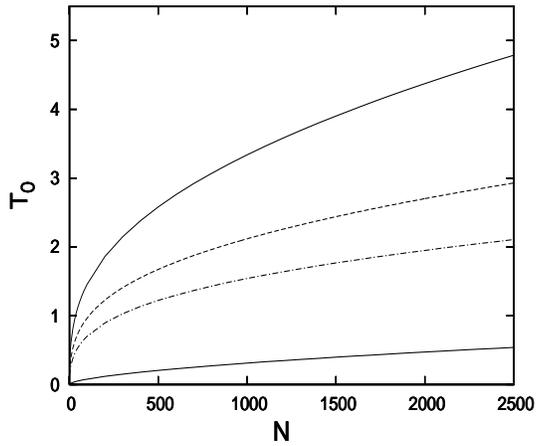}}}
\vspace*{0.5cm}
\caption[]{Bosons number dependence of T$_{0}$ for bosons in a cubic
$50 \times 50 \times 50$ lattice
with $k$ = 0.05 (top), $k$ = 0.02 (dashed),
$k$ = 0.01 (dash-dot), and $k$ = 0 (bottom).}
\end{figure}
\begin{figure}
\resizebox*{3.1in}{2.5in}{\rotatebox{270}{\includegraphics{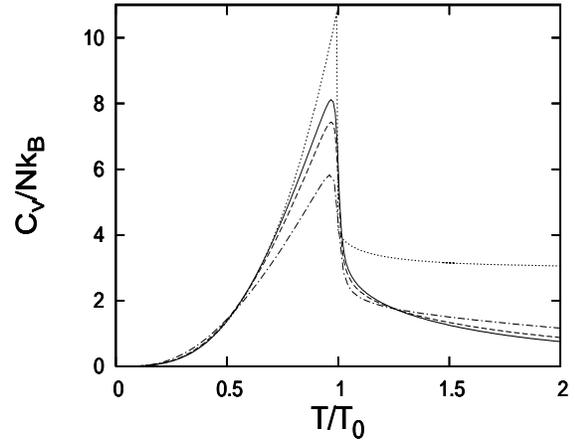}}}
\caption[]{The effect of harmonic potential strength on the specific
heat per boson of 2500 bosons in a  $50 \times 50 \times 50$ cubic lattice.
The results shown are for $k$ = 0.01 (solid line), $k$ = 0.02 (dashed line),
and $k$ = 0.05 (dash-dot line). Dotted line with a sharp peak
is for 2500 bosons in a 3d harmonic trap.
}
\end{figure}
\begin{figure}
\resizebox*{3.1in}{2.5in}{\rotatebox{270}{\includegraphics{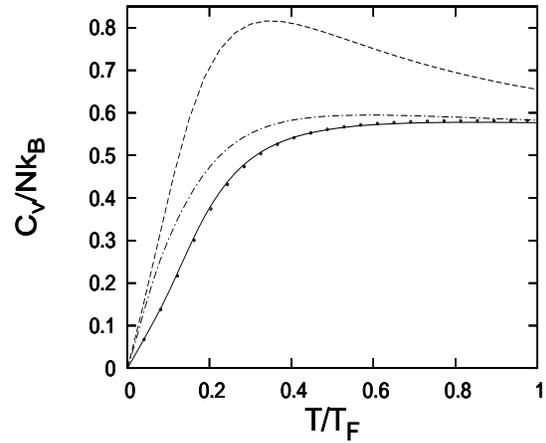}}}
\caption[]{Specific heat per fermion in a 1d optical lattice (of size 3000 sites)
in a harmonic potential with $k$ = 0.0001 (dashed line), 0.0005 (dash-dot),
0.01 (solid line). The filled circles are results in the absence of 
harmonic potential (i.e., for $k$ = 0). The number of fermions ($N$)
considered is 150. $T_F$ is the Fermi temperature.
}
\end{figure}
\begin{figure}
\resizebox*{3.1in}{2.5in}{\rotatebox{270}{\includegraphics{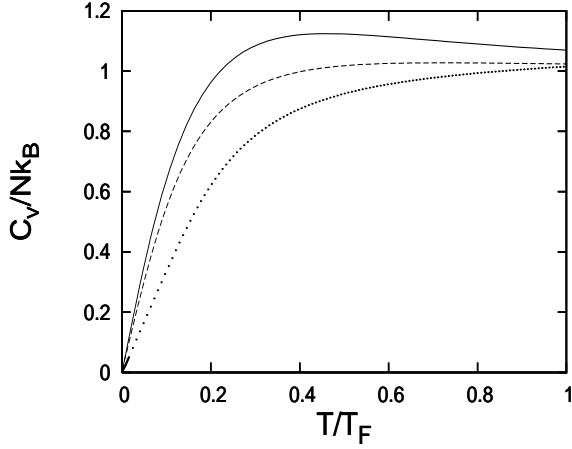}}}
\caption[]{Specific heat per fermion in a 2d optical lattice 
(of size $250 \times 250$) in a harmonic potential with 
$k$ = 0.01 (solid line), and 0.02 (dashed line).
The dots are results in the absence of
harmonic potential (i.e., for $k$ = 0).
The number of fermions considered is 600.
}
\end{figure}
\begin{figure}
\resizebox*{3.1in}{2.5in}{\rotatebox{270}{\includegraphics{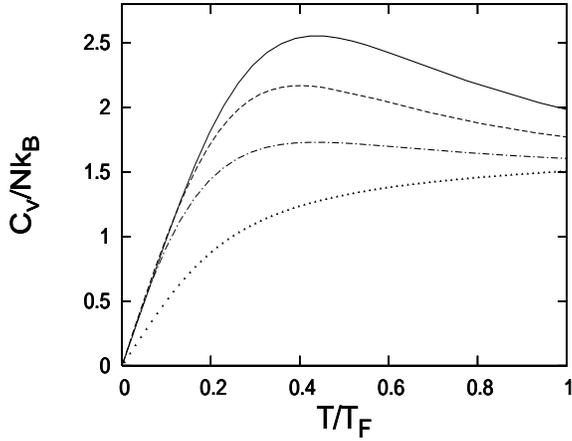}}}
\caption[]{Specific heat per fermion in a 3d optical lattice
(of size $100 \times 100 \times 100$) in a harmonic potential with
$k$ = 0.01 (solid line), 0.02 (dashed line), and 0.05 (dash-dot line).
The dots are results in the absence of
harmonic potential (i.e., for $k$ = 0).
The number of fermions considered is 1000.
}
\end{figure}
\end{document}